# Small-angle scattering reveals heterogeneous nanoparticle size distributions through Bayesian inference

Songhao Yin[†1,2], Yushiro Mori[†1,3], Yui Hayashi[†3], Zuolin Meng[1,4], Lester Canque Geonzon[1,5], Hiroyoshi Nakano[1], Hiroshi Noguchi[1], Masato Okada[3], Koichi Mayumi*[1], and Kumiko Hayashi*[1]

[1]The Institute for Solid State Physics, The University of Tokyo, 5-1-5 Kashiwa-no-ha, Kashiwa, Chiba 277-8581 Japan

[2]Department of Physics, Graduate School of Science, Chiba University, 1-33 Yayoicho, Inage-ku, Chiba, 263-8522 Japan

[3]Department of Complexity Science and Engineering, Graduate School of Frontier Sciences, The University of Tokyo, 5-1-5 Kashiwa-no-ha, Kashiwa, Chiba 277-8561 Japan

[4]Department of Advanced Materials Science, Graduate School of Frontier Sciences, The University of Tokyo, 5-1-5 Kashiwa-no-ha, Kashiwa, Chiba 277-8561 Japan

[5]Institute for Life and Environmental Sciences, University of Tsukuba, 1-1-1 Tennodai, Tsukuba, Ibaraki 305-8577 Japan

†Equal contribution

*Correspondence to Kumiko Hayashi, Koichi Mayumi

hayashi@issp.u-tokyo.ac.jp, kmayumi@issp.u-tokyo.ac.jp

## ABSTRACT

Nanoparticles have attracted considerable attention owing to their broad range of applications. Surface functionalization further expands their functionality, making them useful in fields such as biomedicine, catalysis, sensing, and environmental remediation. In such applications, nanoparticles may form clusters through aggregation under high ionic-strength conditions, while biologically derived nanoparticles can be intrinsically heterogeneous in size. In this study, we apply small-angle neutron scattering (SANS) and small-angle X-ray scattering (SAXS) to nanoparticles and use Bayesian inference to analyze their heterogeneous particle-size distributions. Because SAXS and SANS probe the wavevector dependence of the static scattering intensity, they can sensitively resolve two distinct particle sizes even when a small fraction (< 10%) of larger particles is mixed with smaller particles. We construct multimodal particle-size distribution models within a Bayesian framework and demonstrate the importance of model selection for accurately estimating heterogeneous particle-size distributions. This approach should be broadly applicable to nanoparticle characterization across a wide range of applications.

Nanoparticles have been widely explored in materials science because of their diverse physicochemical properties and broad range of applications. In particular, their use in biological and biomedical fields has expanded rapidly. Functional nanoparticles modified with organic molecules, including antibodies, DNA, and peptides, have been investigated for applications such as biosensing, intracellular delivery, drug delivery, and bioimaging[1, 2, 3, 4, 5].

Small-angle neutron scattering (SANS) offers distinct advantages for characterizing the structures of such nanoparticles[6, 7, 8]. In contrast to X-ray scattering, which depends on electron-density contrast, neutron scattering is sensitive to light elements, such as H, C, and O, and particularly to differences between hydrogen and deuterium. This sensitivity makes SANS particularly well suited for probing surface-modified nanoparticles containing organic components, including those designed for biological and biomedical applications.

Characterization under increased ionic strength is also important for nanoparticles intended for biological applications, because intracellular environments typically have physiologically relevant ionic strengths on the order of 100 mM. Under such conditions, electrostatic screening can promote nanoparticle aggregation, resulting in the coexistence of the original nanoparticles and larger clusters formed by aggregation[1]. Consequently, nanoparticle systems under biologically relevant conditions can exhibit heterogeneous particle-size distributions rather than a single characteristic size.

Accurate characterization of such heterogeneous distributions is therefore essential for understanding the behavior of nanoparticles under high-salt conditions relevant to biological and chemical engineering applications. SANS and Small-angle X-ray scattering (SAXS) are particularly suited to this purpose because their scattering profiles contain information over a broad range of length scales. However, extracting multiple particle-size populations from scattering data requires an appropriate statistical model capable of distinguishing competing distribution models. Conventionally, the structural parameters of nanoparticles are obtained by manually adjusting the model to achieve the best fit to the SAS data or by performing nonlinear least-squares fitting[9]. However, a previous study has pointed out that these fitting procedures are heuristic in nature or tend to converge to local minima in the parameter space[10, 11]. Bayesian inference provides a systematic framework for estimating particle-size distributions and quantitatively comparing models with different numbers of particle populations. In this study, we combine SANS and SAXS with Bayesian model selection to characterize nanoparticles with heterogeneous particle-size distributions and to determine when multiple size populations can be reliably resolved. By applying Bayesian inference, SANS data exhibiting bimodal size distributions, as reported in previous studies, could be analyzed more quantitatively and accurately[12].

In this study, we applied the innovative parameter-fitting approach based on Bayesian inference developed by the previous studies [10, 11] to real experimental SANS and SAXS data. As a simple test case, we first examined dilute dispersions of silica nanoparticles in aqueous $D_2O$ solutions, which represent one of the simplest model systems for nanoparticle scattering. We then extended the analysis to silica nanoparticle dispersions containing 100 mM NaCl. Under increased ionic strength, electrostatic screening alters the interactions between particles, and the contribution of nanoparticle aggregation may become non-negligible[13]. We therefore used Bayesian inference to compare models with different particle-size distributions and to estimate the structural parameters from the experimental data. To validate the inferred particle-size distribution, we additionally performed SAXS measurements on a controlled mixture in which 10% of larger silica nanoparticles were added to a dispersion of smaller nanoparticles. The Bayesian analysis successfully captured the contribution of the larger-particle population, supporting the validity of the inferred size distribution.

Although this study analyzes SANS and SAXS data for silica nanoparticles with a simple structure, the parameter estimation based on Bayesian inference is expected to provide a foundation for future structural analyses of such functional nanoparticles. Additionally, to promote transparency and reproducibility, the Bayesian inference program used in this study will be made publicly available on GitHub. This open-source implementation is expected to facilitate the application of the present approach to a broader range of nanoparticle dispersions[14].

## Results

### Silica nanoparticles dispersed at 1 vol% in $D_2O$ solution

The experimentally measured scattering intensity $I(q)$ for silica nanoparticles at 1 vol% in $D_2O$ (0 mM NaCl)is plotted in Fig. 1a. At this volume fraction, interactions among the silica nanoparticles were expected to be negligible, and the contribution of the structure factor to $I(q)$ was therefore considered to be small[15, 16]. Then the experimental data were ideally fitted using the function $I(q;\theta)$ given in Eq. (11), where the fitting parameters $\theta$ $(= \{\bar{R}, s, \sigma, b\})$ were estimated in Fig. 1b by Bayesian inference based on the previous studies[10, 11], as described in the Methods section. Specifically, Bayesian inference was performed using the exchange Monte Carlo (MC) method[17, 18]. In particular, $\bar{R}$ and $\sigma$ represent the mean radius and standard deviation, respectively, of the particle-size distribution $P(R)$ defined in Eq. (9). The maximum a posteriori (MAP) estimates obtained from the posterior distributions were used for the fit shown in Fig. 1b. The fact that $\bar{R}$ =106.5 Å (Fig. 1b, top, left) consistent with the manufacturer-specified particle size (radius) of 110 Å. Figure 1c shows the particle-size distribution defined by Eq. (9), calculated using the estimated values of $\bar{R}$ and $\sigma$ (Fig. 1b). The shaded region indicates the 95% credible band. In practical nanoparticle synthesis, the particles are not perfectly uniform in size but exhibit a finite size distribution. Therefore, characterizing the overall shape of the particle-size distribution is of practical importance for chemical engineering applications. The resulting unimodal particle-size distribution provides a reference for evaluating the effects of increased ionic strength and nanoparticle aggregation in the following section.

Finally, it is important to confirm that our Bayesian inference worked well. Figure 1d shows the trajectory of the error function $E(\theta)$ obtained by the exchange MC method, where $\theta = \theta_L$, of the $L$-th replica at $\beta = 1$. After the transient time (~5000 MC steps), $E(\theta)$ reaches an equilibrium. To exclude transient effects, the first 10,000 MC steps were discarded as burn-in. After $E(\theta)$ had sufficiently relaxed to an equilibrium, the posterior distributions of $\theta$ were calculated from the time trajectory of $\theta_L$ of the $L$-th replica at $\beta = 1$ over time steps 10001-20000 (Fig. 1b). To assess the adequacy of the exchange MC simulation, the exchange acceptance rates between adjacent replicas were examined, as shown in Fig. 1e. The exchange rate was defined as the fraction of accepted exchanges among the attempted exchanges between adjacent replicas ($l$) and ($l + 1$) after equilibration. The exchange rates exceeded 60% for all adjacent replica pairs, indicating frequent exchanges between replicas. These results support the adequacy of the selected hyperparameters, including the inverse-temperature setting ($\beta_l = 0$ (if $l = 1$) or $\beta_l = 2.2^{l-L}$ (otherwise)) and the number of replicas ($L = 40$). These parameter settings were chosen based on previous studies[10, 11].

**Silica nanoparticles dispersed at 1 vol% in NaCl 100 mM solution**

Figure 2a shows the measured $I(q)$ of the silica nanoparticles at high salt concentration (100 mM NaCl) (orange), compared with the case of 0 mM NaCl (blue). The two curves overlap in the range of $q > 10^{-2}$, which corresponds to the size of individual silica particles (110 Å), whereas a clear difference is observed in the low-$q$ region (the yellow area in Fig. 2a). The aggregation of nanoparticles expected under the high-salt conditions may be reflected in the $q$-dependence of the scattering intensity, $I(q)$, because $q$ corresponds to a characteristic length scale in real space. Based on this observation, we expected that, under the high-salt condition, the particles aggregate, giving rise to a second effective radius corresponding to the cluster size of the aggregates, and that the particle-radius distribution becomes multi-modal (Eqs. (25), (26)).

In Fig. 2b, the scattering data of the silica nanoparticles at a high salt concentration ([NaCl] = 100 mM) were fitted by Bayesian inference using different types of particle-size distribution models. Model 1 is a unimodal particle-size distribution model described by a Gaussian distribution, as defined in Eq. (25) ($K$=1). Model 2 is a bimodal particle-size distribution model described by two Gaussian distributions, as defined in Eq. (25) ($K$=2). Model 3 is a bimodal particle-size distribution model described by two Gamma distributions, as defined in Eq. (26) ($K$=2). The posterior distributions obtained from Bayesian inference using these models are provided in the Supplementary Information (Figs. S1-S3). Model 1, which assumes a unimodal particle-size distribution, did not adequately fit the experimental data, particularly in the low-$q$ region, whereas Models 2 and 3, which assume bimodal distributions, generally reproduced the data well in this region. The Bayesian free energy ($F$) and marginal likelihood $p(D|\text{Model } k)$ indicate that Model 2 was selected, as it had the lowest Bayesian free energy and the highest marginal likelihood (Fig. 2b, bottom).

For Model 2, the obtained particle size distribution $P(R)$ is shown in Fig. 2c. The first peak in the particle-size distribution was located at nearly the same radius as that of the silica particles (blue) in the absence of NaCl shown in Fig. 1a. This observation suggests that the scattering intensity observed for the sample containing 100 mM NaCl in Fig. 2a can be interpreted as arising from the emergence of a second effective radius (an orange arrow depicted in Fig. 2c) due to aggregation, in addition to the dispersed silica particles at a dilute concentration. As indicated by the arrow in Fig. 2c, even when clusters formed by aggregation accounted for only a few percent of the sample, SANS measurements of $I(q)$ may have sensitively detected these larger structures. Since $I(0)$ is proportional to the sixth power of the particle radius, $I(q)$ is highly sensitive to the presence of even a small population of larger particles. We will discuss this point further in the Discussion section.

**Dynamic Light Scattering (DLS) measurements**

Figure 2d shows the DLS-derived hydrodynamic radius distributions of silica nanoparticles at NaCl concentrations of 0 mM and 100 mM. Unlike the $I(q)$ profiles (Fig. 2a), DLS was found to detect the aggregation of silica nanoparticles at 100 mM NaCl primarily as an increase in the average particle size, as observed in both the hydrodynamic radius distribution (Fig. 2d, left) and the correlation function $g^{(2)}(\tau) - 1$ (Fig. 2d, right) (Eq. (4)). The DLS correlation function $g^{(2)}(\tau) - 1$ shows a slower relaxation for silica nanoparticles in 100 mM NaCl than for those in 0 mM NaCl. The DLS results did not clearly reveal the emergence of a distinct aggregated population in addition to the original monodisperse silica nanoparticle population, as observed in the particle-size distribution $P(R)$ (Fig. 2c). Even from a comparison of $g^{(2)}(\tau) - 1$, we could only determine that the average value of $\Gamma$ decreased at 100 mM NaCl (Fig. 2d, right panel, orange curve).

**Validation of the analysis using mixed-size silica particles**

We interpreted the behavior of $I(q)$ for silica nanoparticles at 100 mM NaCl in Fig. 2a in terms of a heterogeneous particle-size distribution. To test this interpretation more directly, we prepared artificial mixtures of two silica nanoparticle populations, NP1 and NP2, with different nominal radii of 70 Å and 110 Å, respectively, at a particle volume fraction of 1 vol%. This approach allowed us to examine whether larger particles could be distinguished within a population of smaller nanoparticles under well-defined mixing conditions. Here, we performed SAXS measurements on this mixture to examine its scattering intensity profiles and to demonstrate that Bayesian inference can be applied to both SANS and SAXS data.

First, Bayesian fitting was performed separately for each nanoparticle type (black lines in Fig. 3a), yielding estimated radii of 79 Å and 139 Å for NP1 and NP2, respectively (see the Supplementary Information (Fig. S4) for the posterior distributions of the parameters). Here we used the model described by Eq. (25) with $K$=2. Then, Figure 3b shows the SAXS intensity profile $I(q)$ for a mixture of NP1 and NP2 at a volume ratio of 9:1, with a total particle volume fraction of 1 vol%. Compared with the NP1-only sample (Fig. 3b, blue), the addition of 10 % NP2 resulted in a pronounced change in the SAXS intensity profile in the low-$q$ region (Fig. 3b, red). The behavior, in which the scattering from smaller particles is reflected in the higher-$q$ region and that from larger particles in the lower-$q$ region is similar to that observed for silica nanoparticle aggregation at high salt concentration (Fig. 2a). Figure 3c shows the result of Bayesian estimation for the NP1–NP2 mixture. From these results, we estimated the radii of NP1 and NP2 and their mixing ratio, as 79 Å, 121 Å and 9:1. These estimates were consistent with the known particle radii and mixing ratio of the prepared sample, confirming the validity of our Bayesian analysis. Figure 3d shows the particle size distribution calculated using the parameters of the bimodal Gaussian model obtained by Bayesian inference.

**DLS characterization of silica nanoparticle mixtures**

Figure 3e shows the DLS results for NP1, NP2, and the NP1–NP2 mixture. In the DLS measurements of the NP1–NP2 mixture (Fig. 3e, right), the larger NP2 particles dominated the correlation function $g^{(2)}(\tau) - 1$ despite the smaller NP1 particles accounting for 90% of the total particle volume. This is because DLS is strongly biased toward larger particles due to their greater scattering intensity, making it difficult to resolve the sizes and relative proportions of the two particle populations. Similarly, the hydrodynamic radius distribution (Fig. 3e, left) only indicates an increase in the average particle size of the mixture, as observed for the sample at 100 mM NaCl (Fig. 2d). In contrast, the scattering intensity $I(q)$ provides structural information over a broad range of $q$, allowing structural contributions from particles of different sizes to be distinguished through their characteristic $q$-dependence (Fig. 3b). These results suggest that applying Bayesian inference to $I(q)$ is well suited for characterizing heterogeneous particle size distributions in this artificially prepared mixture of silica nanoparticles of different sizes.

**Discussion**

In this study, we demonstrated that Bayesian analysis of SANS and SAXS intensity profiles enables the identification of heterogeneous particle-size distributions in nanoparticle dispersions. In particular, our analysis distinguished two nanoparticle populations even when the fraction of larger particles was as small as 10% by volume. This capability was demonstrated using both salt-induced aggregation of silica nanoparticles and artificially prepared mixtures of nanoparticles with different sizes. The results highlight the importance of appropriate particle-size distribution models and Bayesian model selection for characterizing heterogeneous nanoparticle systems. The quantitative characterization of heterogeneous particle-size distributions is also important for biological nanoparticles, such as extracellular vesicles (EVs)[19, 20], including exosomes[21]. Because EVs exhibit substantial heterogeneity in size and composition, resolving distinct particle populations and estimating their relative proportions may provide valuable information for understanding their biological functions and evaluating sample quality. Blood- and blood cell-derived colloidal materials also exhibit heterogeneous particle size distributions[22]. The Bayesian approach developed in this study may therefore provide a useful framework for analyzing heterogeneous biological nanoparticle dispersions, in the future.

A key advantage of Bayesian analysis of SANS and SAXS intensity profiles over conventional DLS analysis is its ability to resolve heterogeneous particle-size distributions. In SANS and SAXS, the scattering intensity $I(q)$ is measured over a broad range of scattering vectors $q$, providing structural information at different length scales. Therefore, when the sample contains particles of different sizes, their contributions to $I(q)$ can be distinguished through Bayesian inference using an appropriate particle-size distribution model. This finding can be explained in terms of numerical simulation of the scattering intensity $I(q)$[9, 23] and the autocorrelation function $g^{(2)}(\tau) - 1$[24], defined in the Method section (Eq. (4)), as follows:

$$I(q) = \phi_1(R_1)^3 P(q, R_1) + \phi_2(R_2)^3 P(q, R_2), \quad (1)$$

$$g^{(2)}(\tau) - 1 = [\omega_1 e^{-\Gamma_1 \tau} + (1 - \omega_1) e^{-\Gamma_2 \tau}]^2, \quad (2)$$

where $\phi_1$ is a relative volume fraction varied from 0 to 1 and $\omega_1 = \phi_1(R_1)^3/(\phi_1(R_1)^3 + \phi_2(R_2)^3)$, and $\phi_1 + \phi_2 = 1$. The model assumes homogeneous monodisperse spheres, identical scattering contrast, negligible interparticle interactions, and hydrodynamic radii equal to the geometric radii. Figure 4a shows the calculated results for $I(q)$, using Eq. (1), and $g^{(2)}(\tau) - 1$, using Eq. (2), for $R_1 = 70$ Å (blue), $R_2 = 110$ Å (green), and a mixture of NP1 and NP2 at a volume ratio of 9:1 (red). The characteristic features of $I(q)$ and $g^{(2)}(\tau) - 1$ for the heterogeneous particle mixtures examined in this study can be clearly observed. $I(q)$ captures the structural contributions of both the larger and smaller particles because their characteristic length scales correspond to different $q$ ranges. In addition, because $I(0)$ scales with the sixth power of the particle radius, the contribution from the larger particles becomes particularly pronounced in the low-$q$ region. These features are consistent with those observed in the SANS/SAXS and DLS measurements shown in Figs. 2 and 3. In practice, commercially purchased silica nanoparticles have a distribution of particle sizes that includes larger particles. Therefore, we also considered a more extreme case with $R_2 = 260$ Å through simulations (Fig. 4b). When particles twice as large are included in the mixture, the differences in $I(q)$ become even more pronounced. In contrast, $g^{(2)}(\tau) - 1$ appears to be dominated by the contribution from larger particles, making it difficult to distinguish between the small and large particle populations in the mixture. By exploiting these properties of $I(q)$, Bayesian inference enables the estimation of $R_1$, $R_2$, and their mixing ratio. Thus, the analysis of $I(q)$ using SANS or SAXS is particularly effective for characterizing samples with heterogeneous particle-size distributions.

As shown by the conventional fitting analysis using SasView[25] in the Supplementary Information (Fig. S5), the scattering data obtained at 100 mM NaCl could not be fully reproduced by the selected fitting model. Without systematic model comparison, however, it is difficult to determine whether such discrepancies arise from an inadequate particle-size distribution model. In contrast, Bayesian inference provides a systematic framework for estimating particle-size distributions and quantitatively comparing competing models based on their Bayesian free energies and marginal likelihoods. By combining SANS and SAXS measurements with Bayesian model selection, we demonstrated that distinct particle-size populations can be identified and their sizes and mixing ratios estimated, even when one population constitutes only a small fraction of the sample.

Finally, in this study, although Bayesian inference often involves computationally demanding methods, including parallel computing, we have developed a simple Python program that allows experimental researchers to perform Bayesian parameter estimation and model selection without requiring advanced computational expertise. To promote accessibility, transparency, and reproducibility, the program will be made publicly available on GitHub[14]. We hope that this implementation will encourage the broader use of Bayesian inference in the analysis of SANS and SAXS data and contribute to the quantitative characterization of heterogeneous nanoparticle systems.

## Methods

### Samples

Silica nanoparticle suspensions (LUDOX TM-50, Sigma-Aldrich, product No. 420778-1L), sodium chloride (NaCl, FUJIFILM Wako, product No. 191-01665), and deuterium oxide ($D_2O$, KANTO CHEMICAL, product No. 10086-23) were used as received. To prepare the NaCl-free silica sample (NaCl 0 mM condition), the silica suspension was diluted with pure water ($H_2O$) and $D_2O$ to a final silica volume fraction of 1 vol%. The volume fraction of $D_2O$ in the solvent mixture was adjusted to 0.9. For the high-salt silica sample (NaCl 100 mM condition), NaCl was first dissolved in $D_2O$ to prepare a stock solution with a mass concentration of approximately 9%. The silica suspension was then diluted with $H_2O$, $D_2O$, and the NaCl/$D_2O$ stock solution to achieve a final silica volume fraction of 1 vol% and a NaCl concentration of 100 mM, while maintaining the $D_2O$ volume fraction in the solvent at 0.9.

### SAXS instrument

SAXS measurements were performed at the BL-10C beamline of the Photon Factory (PF), High Energy Accelerator Research Organization (KEK), Japan. The incident X-ray wavelength was 1.54 Å, and the scattering patterns were recorded using a PILATUS3 2M detector (Dectris, Switzerland) with a a sample-to-detector distance of 1.32 m. The beam size at the sample position was 0.167 mm (vertical) × 0.315 mm (horizontal), and each scattering pattern was collected with an exposure time of 420 s at room temperature. The beam center and sample-to-detector distance were calibrated using silver behenate. A glassy carbon standard was measured to convert the scattering intensities to absolute units. Data reduction and conversion of the two-dimensional scattering patterns into one-dimensional profiles were performed using the Red2D package in Igor Pro. The scattering intensities were corrected for exposure time, sample thickness, and transmittance, followed by subtraction of the corresponding background scattering. The corrected patterns were circularly averaged to obtain $I(q)$ as a function of the scattering vector $q$.

### SANS instrument

SANS experiments were performed on the SANS-U instrument at the Institute for Solid State Physics, The University of Tokyo, which is installed at the JRR-3 research reactor of the Japan Atomic Energy Agency in Tokai, Japan. A neutron wavelength of 7.0 Å with a wavelength spread ($\Delta\lambda/\lambda$) of 10% was employed. Measurements were conducted at three sample-to-detector distances (SDDs) of 1, 4, and 12 m to cover a wide $q$ range. All experiments were carried out at 298 K. Data reduction followed standard calibration procedures. Detector sensitivity correction was performed using scattering from a Lupolen standard, air, and $B_4C$ measured at an SDD of 2 m. Background scattering from air, the empty cell, and $B_4C$ was measured separately at SDDs of 12, 4, and 1 m, using exposure times identical to those of the corresponding sample measurements. These background datasets were subtracted from the raw sample scattering intensities. The final scattering intensity was converted to absolute scale using the data reduction package within Igor Pro. The reduced scattering intensities were circularly averaged, and the averaged scattering intensity $I(q)$ was plotted as a function of the magnitude of the scattering vector $q$. The error bars for $I(q)$ were given by $\Delta I = \pm\epsilon$, where $\epsilon$ represents the standard deviation of the circular averaging.

**Dynamic light scattering experiment**

To characterize the hydrodynamic size and dispersion state of the silica nanoparticles, dynamic light scattering (DLS) measurements were performed. The DLS experiments were carried out using a DLS/SLS-5000 compact goniometer (ALV, Langen, Germany) equipped with a 22 mW He-Ne laser operating at a wavelength of $\lambda$=632.8 nm. All measurements were conducted at a fixed scattering angle of $\Theta$=90° and a controlled temperature of 25 ℃.

The time-intensity autocorrelation functions of the scattered light were recorded and subsequently analyzed using the CONTIN algorithm to obtain the distribution function of the relaxation rate $\Gamma$. The magnitude of the scattering vector $q_{\mathrm{DLS}}$ is given by $q_{\mathrm{DLS}} = \frac{4\pi n_r}{\lambda}\sin\left(\frac{\Theta}{2}\right)$, where $n_r$ is the refractive index of the solvent at the measurement temperature. The hydrodynamic radius $R_{\mathrm{h}}$ of the nanoparticles was then estimated from $\Gamma$ using the Stokes-Einstein equation: $R_{\mathrm{h}} = \frac{k_{\mathrm{B}}T{q_{\mathrm{DLS}}}^2}{6\pi\eta\Gamma}$, where $k_{\mathrm{B}}$ is the Boltzmann constant, $T$ is the absolute temperature, and $\eta$ is the viscosity of the solvent.

The first-order correlation function, $g^{(1)}(\tau)$, is the autocorrelation function of the scattered electric field,

$$g^{(1)}(\tau) = \frac{\langle E^*(q,t)E(q,t+\tau)\rangle}{\langle I(q,t)\rangle}, \tag{3}$$

where $E^*$ is the complex conjugate of $E$. For a dispersion containing a single type of particle, $g^{(1)}(\tau)$ follows the relation $g^{(1)}(\tau) = e^{-Dq^2\tau}$. Then, the second-order correlation function, $g^{(2)}(\tau)$, is defined by

$$g^{(2)}(\tau) = \frac{\langle I(q,t)I(q,t+\tau)\rangle}{\langle I(q,t)\rangle^2}. \tag{4}$$

For a dispersion containing a single type of particle, $g^{(2)}(\tau)$ follows the relation $g^{(2)}(\tau) = 1 + e^{-2Dq^2\tau}$. $g^{(2)}(\tau) - 1 = \left|g^{(1)}(\tau)\right|^2$. From this equation, $g^{(2)}(\tau) - 1 = e^{-2\Gamma\tau}$.

**Scattering intensity in the absence of a structure factor**

When $\boldsymbol{r}$ and $\boldsymbol{q}$ are defined as the position vector and scattering vector, respectively, the scattering amplitude $F(\boldsymbol{q})$ is represented by

$$F(\boldsymbol{q}) = \int_V \rho(\boldsymbol{r})e^{-i\boldsymbol{q}\cdot\boldsymbol{r}}\,d\boldsymbol{r}, \tag{5}$$

where $\rho(\boldsymbol{r})$ is the scattering length density of sample (nanoparticles) and $V$ is the volume of the nanoparticle.

In the case of scattering by spherically symmetric particles, Eq. (5) can be rewritten as

$$F(q) = 4\pi(\rho_{\mathrm{s}} - \rho_{\mathrm{m}}) \int_0^R \frac{\sin(qr)}{qr} r^2 dr, \qquad (6)$$

where $\rho_{\mathrm{s}}$ and $\rho_{\mathrm{m}}$ are the uniform scattering length densities of the nanoparticle and medium ($D_2O$), respectively, and $R$ is the radius of the nanoparticle. Here, $r$ and $q$ represent the magnitudes of $\boldsymbol{r}$ and $\boldsymbol{q}$, respectively. Using relations $\Delta\rho = \rho_{\mathrm{s}} - \rho_{\mathrm{m}}$ and $V = 4\pi R^3/3$,

$$F(q, R) = \Delta\rho V(R)\Phi(qR), \qquad (7)$$

where

$$\Phi(qR) = \frac{3[\sin(qR) - qR\cos(qR)]}{(qR)^3}. \qquad (8)$$

Note that $\Phi(0) = 1$. In real experiments, the radius $(R)$ of nanoparticles is distributed around an average value $(\bar{R})$. This distribution is assumed to be approximated by a Gaussian function:

$$P(R) = \frac{1}{\sqrt{2\pi}\sigma} e^{-(R-\bar{R})^2/2\sigma^2}. \qquad (9)$$

Using Eq. (9), $\bar{V} = 4\pi/3\overline{R^3} = 4\pi\bar{R}^3(1 + 3\sigma^2/\bar{R}^2)/3$ and the number of particles per unit volume $N$ is defined by $N = \phi/\bar{V}$ and $\phi$ is the volume fraction of the sample. Then, $I(q)$ is written as

$$I(q) = N \int_0^\infty F(q)^2\, P(R)\, dR. \qquad (10)$$

In real experiments, the background noise $(b)$ and measurement time of the experiment $(t)$ contribute to $I(q)$. Furthermore, because the measured scattering intensity in SANS is proportional to the measurement time $t$, the data used in this study were normalized by $t$. Therefore, the expression can be rewritten as follows.

$$\frac{I(q)}{t} = s \int_0^\infty [V(R)\Phi(qR)]^2\, P(R) dR + b, \qquad (11)$$

where $s$ is the scale factor defined as $s = N(\Delta\rho)^2$. Hereafter, the normalized scattering intensity $I(q)/t$ is expressed as $I(q;\theta)$ to explicitly indicate its dependence on the parameter set $\theta = \{\bar{R}, s, \sigma, b\}$.

**Bayesian inference**

After the scattering intensities of the silica nanoparticles were measured in the SANS experiment, the structural parameters $\theta = \{\bar{R}, s, \sigma, b\}$ of the nanoparticles in Eq. (11) were estimated using Bayesian inference. The framework of Bayesian inference was constructed based on the references [10, 11, 17], as outlined below.

The experimental data of the scattering intensity $(q_i, I_i)$ are denoted as $(x_i, y_i)$, where $i$ represents the number of data points $(i = 1, \ldots, n)$. We consider the noise model that

$$y_i = I(x_i; \theta) + \epsilon_i. \qquad (12)$$

When the noise $\epsilon_i$ is a Gaussian with zero mean and variance ${\sigma_{\mathrm{g},i}}^2$, the likelihood of the observation can be written as follows:

$$p(y_i|x_i,\theta) = \frac{1}{\sqrt{2\pi{\sigma_{\mathrm{g},i}}^2}} e^{-\left(y_i - I(x_i;\theta)\right)^2/2{\sigma_{\mathrm{g},i}}^2}. \quad (13)$$

The values of $\{\sigma_{\mathrm{g},i}\}$ $(i = 1, \dots, n)$ were experimentally determined from the measurement errors as $\epsilon$ in Sec. 1.2.

In the framework of Bayesian inference, the posterior density $p(\theta|D)$ of parameter $\theta$ given a data set $D$ $(= \{(q_i, I_i)\}$ $(i = 1, \dots, n))$ is expressed by using Bayes' theorem:

$$p(\theta|D) = \frac{p(D|\theta)p(\theta)}{p(D)}. \quad (14)$$

In this study, $p(D|\theta) = \prod_{i=1}^{n} p(y_i|x_i,\theta)$ where $p(y_i|x_i,\theta)$ is described by Eq. (13), $p(\theta) = \varphi(\theta)$ ($\varphi(\theta)$: the prior density), and $p(D)$ is constant for a given data set. Finally, we obtain

$$p(\theta|D) = \frac{1}{Z} e^{-E(\theta)/2} \varphi(\theta), \quad (15)$$

Where the mean square error function is described as

$$E(\theta) = \sum_{i=1}^{n} \left[ \frac{(y_i - I(x_i;\theta))^2}{\sigma_{\mathrm{g},i}^2} + \ln\left(2\pi\sigma_{\mathrm{g},i}^2\right) \right], \quad (16)$$

and the normalization constant $Z$ is defined as

$$Z = \int e^{-E(\theta)/2} \varphi(\theta) d\theta. \quad (17)$$

The prior distribution $\varphi(\theta) = p(\bar{R})p(s)p(\sigma)p(b)$. When $\mathrm{Gamma}(x;\alpha,\gamma) = \frac{e^{-\frac{x}{\gamma}}}{\Gamma(\mathrm{k})\gamma^{\alpha}} x^{\alpha-1}$, $p(\bar{R})$, $p(\sigma)$, $p(b)$ and $p(s)$ are described by

$$p(\bar{R}) = \text{const.} \quad \text{for } [5:500], \quad (18)$$

$$p(\sigma) = \text{const.} \quad \text{for } [0.01:100], \quad (19)$$

$$p(b) = \text{const.} \quad \text{for } [0:0.5], \quad (20)$$

$$p(s) = \mathrm{Gamma}(s;\alpha_s,\gamma_s), \quad (21)$$

where $\alpha_s = 8.0$ and $\gamma_s = 0.4$. These parameters were determined based on empirical values.

The parameter set $\theta^*$ that realized the maximum posterior density $p(\theta^*|D)$ is obtained by using the following replica-exchange Markov Chain Monte Carlo (MCMC) method [18]. The replica-exchange MCMC method simultaneously simulates $L$ samplings with different inverse temperatures $\{\beta_l : l = 1, \dots, L\}$, where $\beta_l = 0$ (if $l = 1$) or $\beta_l = 2.2^{l-L}$ (otherwise) with $L = 40$. Because $Z$ is a constant in Eq. (15), in the simulation, we consider realizing the maximization of the following quantity $q(\theta_l;\beta_l)$, instead of $p(\theta|D)$ using the replica-exchange MCMC method,

$$q(\theta_l;\beta_l) \propto e^{-\beta E(\theta_l)/2} \varphi(\theta_l). \quad (22)$$

Then, the target density to maximize in the exchange MC simulation is described as

$$p(\theta_1, \dots, \theta_L) = \prod_{l=1}^{L} q(\theta_l; \beta_l). \quad (23)$$

In the simulation, $\theta_l$ is updated as $\theta_l'$ based on the probability $\min\big(1, q(\theta_l'; \beta_l)/q(\theta_l; \beta_l)\big)$ by a conventional MCMC method with the Metropolis algorithm (it is performed for all $\{\theta_l : l = 1, \dots, L\}$) [26, 27].

Then the parameters between adjacent replicas $\theta_l$ and $\theta_{l+1}$ are exchanged at each step with the probability $u = \min(1, v)$ where

$$v = \frac{q(\theta_{l+1}; \beta_l) q(\theta_l; \beta_{l+1})}{q(\theta_l; \beta_l) q(\theta_{l+1}; \beta_{l+1})} = e^{(\beta_{l+1} - \beta_l)(E(\theta_{l+1}) - E(\theta_l))/2}. \quad (24)$$

By repeating these updates, the trajectories of $\{\theta_l : l = 1, \dots, L\}$ are obtained that maximize the target density (Eq. (23)). Then the posterior density of each parameters can be calculated from the trajectory of $\theta_L$ where $\beta_L = 1$. The posterior distributions of the structural parameters ($\theta = \{\bar{R}, \sigma, b\}$) and the MAP estimates were calculated from the trajectory of $\theta_L$.

## 2.6 Comparison of scattering models

Under high-salt conditions, the particle size distribution of silica nanoparticles is expected to include not only radii corresponding to individual particles, but also components with larger effective radii associated with the formation of aggregates.

$$P(R) = \sum_{k=1}^{K} a_k \frac{1}{\sqrt{2\pi}\sigma_k} e^{-(R - \bar{R}_k)^2 / 2\sigma_k^2}, \quad (25)$$

where $\sum_{k=1}^{K} a_k = 1$. For the gamma distribution,

$$P(R) = \sum_{k=1}^{K} a_k \frac{1}{\Gamma(\alpha_k)\theta_k^{\alpha_k}} R^{\alpha_k - 1} e^{-R/\theta_k}, \quad (26)$$

where $\sum_{k=1}^{K} a_k = 1$.

The posterior distribution of parameter $K$ given the data was obtained using $p(\theta|D)$ in Eq. (14).

$$p(\text{Model } K|D) = \frac{\int p(D, \theta, \text{Model } K) d\theta}{\sum_K \int p(D, \theta, \text{Model } K) d\theta},$$

$$= \frac{e^{-F(\text{Model } K)} \varphi(\text{Model } K)}{\sum_K e^{-F(\text{Model } K)} \varphi(\text{Model } K)}, \quad (27)$$

where

$$F(K) = -\log Z(K). \quad (28)$$

## Acknowledgment

This neutron experiment at SANS-U[C1-2] was carried out by the JRR-3 general user program managed by the Institute for Solid State Physics, the University of Tokyo (proposal No. 25535). This work was supported by JST FOREST Program (grant number JPMJFR2120) and Data Creation and Utilization-Type Material Research and Development Project grant number JPMXP1122714694.

**Author contributions**

K.H. and K.M. conceived and planned the study, with input from Y.H. and M.O. S.Y. and Y.M. analyzed the data using Bayesian inference, with assistance from Y.H., H.N. and H.N. Z.M., L.C.G., K.H., and K.M. performed the experiments. Z.M. performed the simulation (Fig. 4).

**Competing interests**

The authors declare no conflicts of interest.

**Data availability**

The data supporting the findings of this study are available from the corresponding authors upon reasonable request.

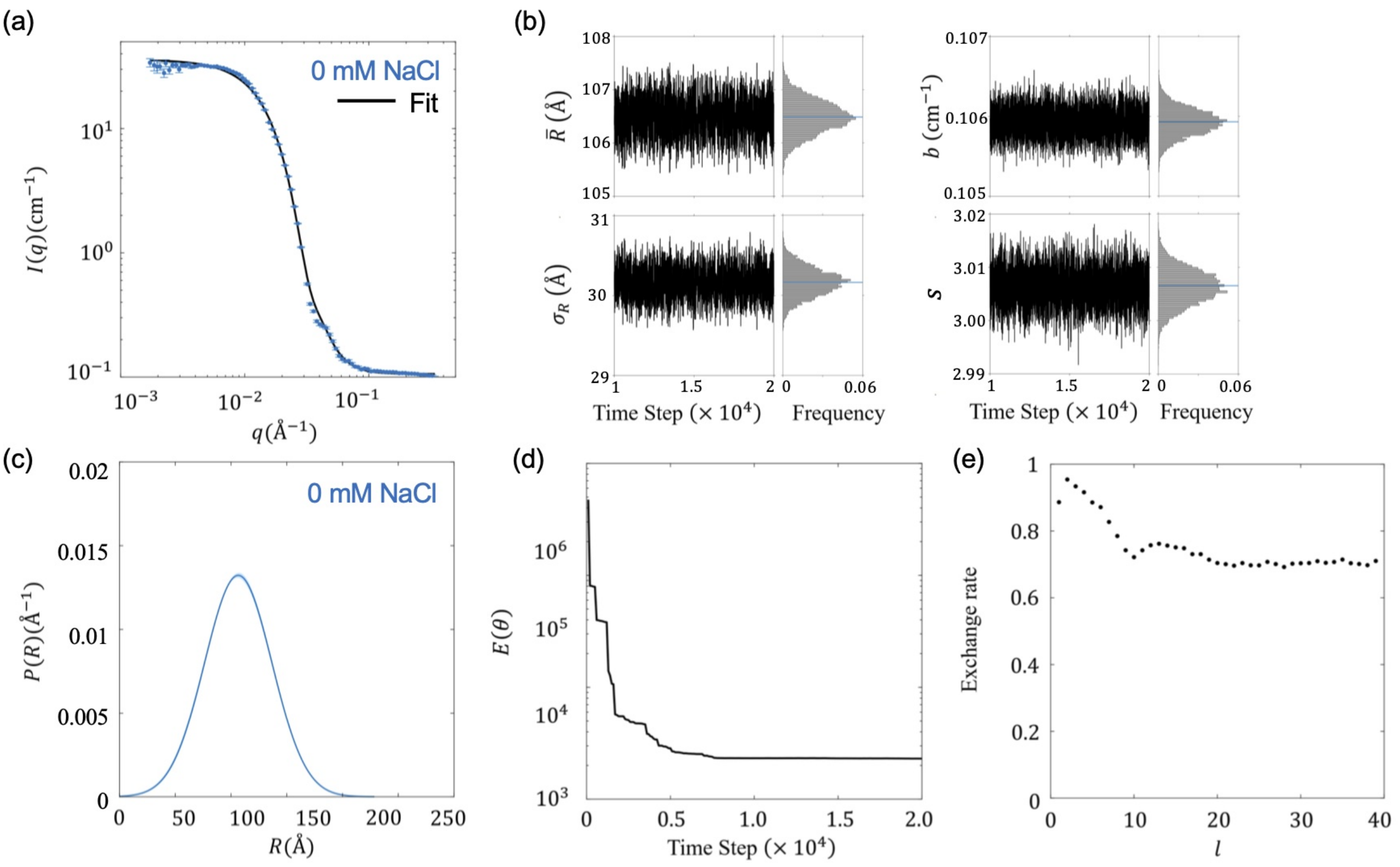

**Fig. 1. Experimentally measured scattering intensity of silica nanoparticles at 1 vol% in $D_2O$ solution (0 mM NaCl). a** The blue circles represent the experimental data, and the black line represents the fitting curve obtained by Bayesian inference based on the scattering model defined in Eq. (11). The error bars indicate the experimental uncertainties. **b** Posterior distributions of the parameters for silica nanoparticles at 1 vol% in $D_2O$ solution. The distributions were obtained from the trajectories of $\theta_L$ $(= \{\bar{R}, s, \sigma, b\})$ using the framework of Bayesian inference (Methods). The scale factor $s$ is dimensionless, with a normalization factor of $(\sqrt{2\pi}/9)N_0 \times 10^{-6}$ (cm$^{-1}$Å$^{-6}$), where $N_0 = 1/(\bar{R}^3 + 3\bar{R}\sigma^2)$. The blue lines indicate the maximum a posteriori (MAP) estimates. **c** Particle-size distribution for the silica nanoparticle, defined in Eq. (9). The parameters $\bar{R}$ and $\sigma$ were estimated by Bayesian inference (Fig. 1b). The shaded region indicates the 95% credible interval. **d** Trajectory of the error function ($E(\theta)$), defined in Eq. (16), during the exchange MCMC calculation, plotted as a function of the MC step. **e** Exchange rate between adjacent replicas, calculated from the number of exchanges between neighboring replicas $l$ and $l + 1$ at equilibrium over 10000 steps.

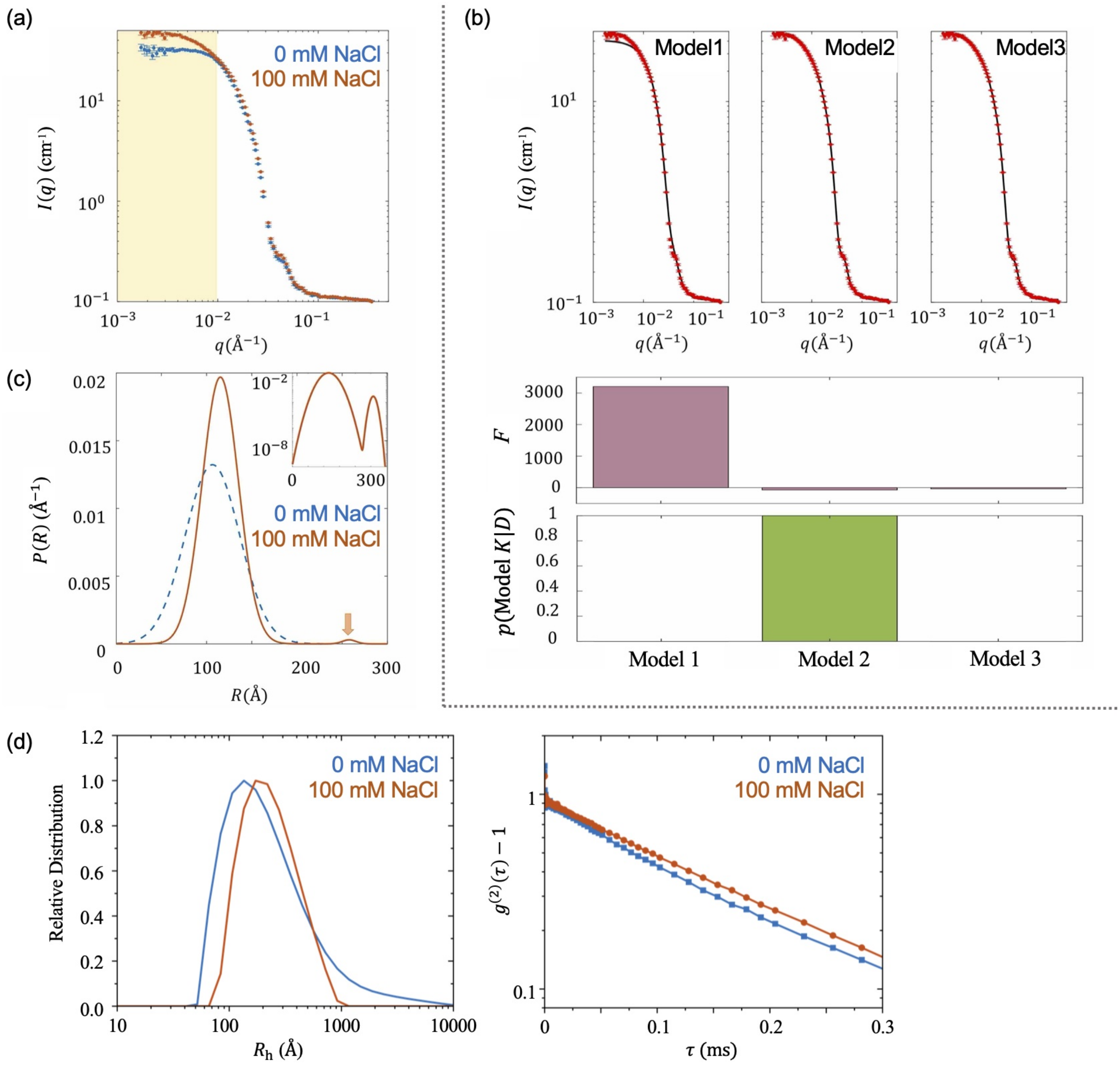

**Fig. 2 Experimentally measured scattering intensity of silica nanoparticles dispersed at 1 vol% in $D_2O$ containing 100 mM NaCl. a** The scattering profile obtained in the presence of 100 mM NaCl (orange) is compared with the NaCl-free profile (blue) shown in Fig. 1. **b** Bayesian fitting of the experimentally measured scattering intensity of silica nanoparticles dispersed at 1 vol% in $D_2O$ containing 100 mM NaCl (top). The three graphs show the fitting results obtained using Models 1 (Eq. (25), $K$=1), Model 2 (Eq. (25), $K$=2), and Model 3 (Eq. (26), $K$=2), respectively. See the Supplementary Material (Figs. S1–S3) for the posterior distributions of the parameters. The Bayesian free energy ($F$) and posterior model probability $p(\text{Model}\, k|D)$ indicate that Model 2 was selected (bottom). **c** The particle-size distribution ($P(R)$) was represented by the two-component model defined in Eq. (25) with $K$=2 (orange). The blue line corresponds to the NaCl-free condition shown in Fig. 1c. **d** DLS-derived hydrodynamic radius ($R_\mathrm{h}$) distributions (left) and correlation functions ($g^{(2)}(\tau)-1$) (Eq. (4)) (right) of silica nanoparticles at 0 mM NaCl (blue) and 100 mM NaCl (orange).

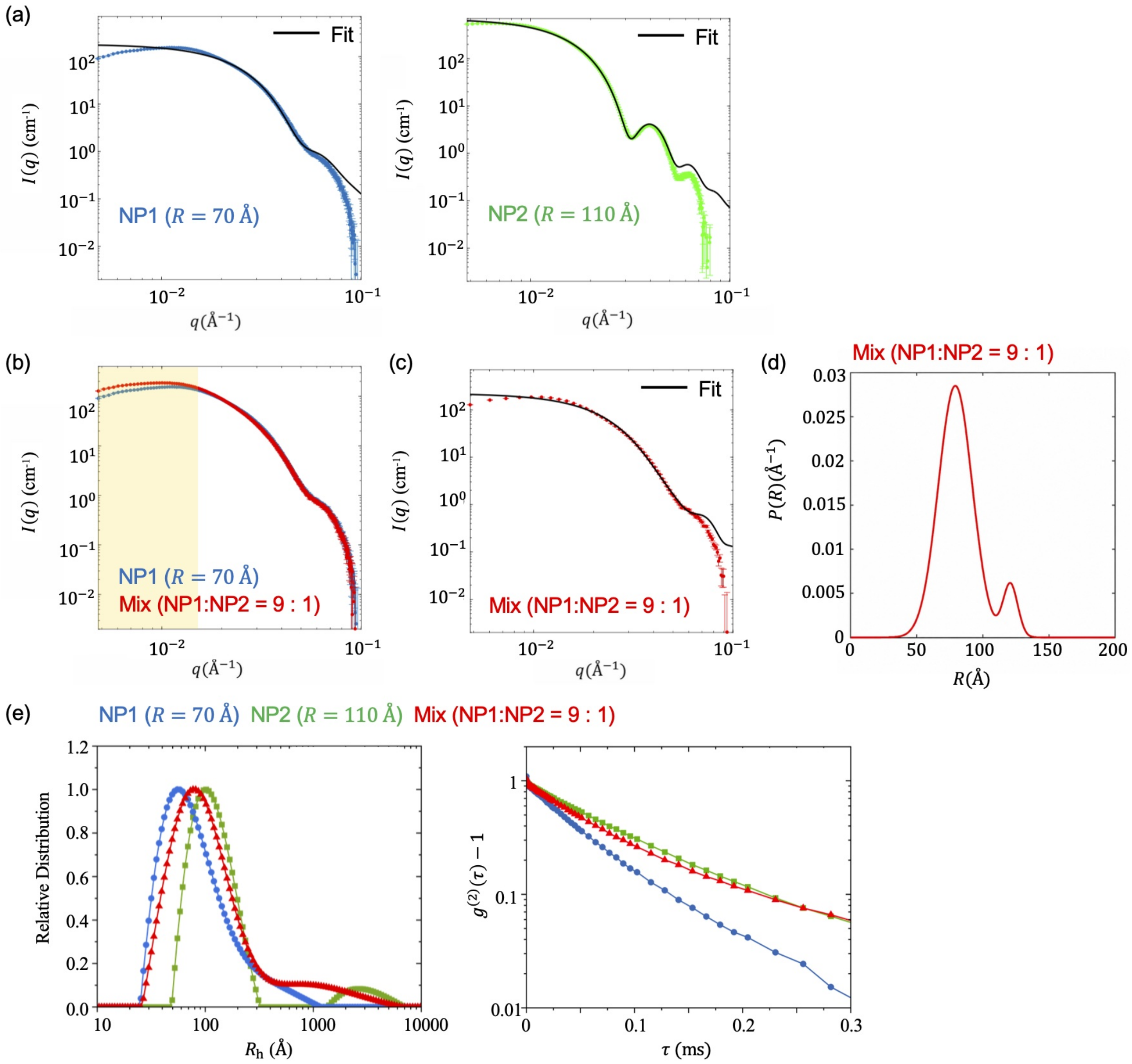

**Fig. 3 Scattering intensity of artificially mixed silica nanoparticle dispersions measured by using SAXS.** **a** SAXS intensities $I(q)$ for silica nanoparticles (NP1) with a nominal radius of 70 Å and those (NP2) with a nominal radius of 110 Å. The black lines represent Bayesian fits to $I(q)$ (See the Supplementary Information (Fig. S4) for the posterior distributions of the parameters). **b** SAXS intensity $I(q)$ for a mixture of NP1 and NP2 at a volume ratio of 9:1. The contribution from the small fraction of larger nanoparticles (NP2) is evident in the low $q$ region. **c** Bayesian fitting result for the SAXS intensity $I(q)$ of the NP1–NP2 mixture using by the two-component model defined in Eq. (25) with $K$=2. **d** The particle-size distribution $P(R)$ for the NP1–NP2 mixture. **e** DLS-derived hydrodynamic radius ($R_\mathrm{h}$) distributions (left) and correlation functions ($g^{(2)}(\tau) - 1$) (Eq. (4)) (right) of silica nanoparticles for NP1, NP2 and the NP1–NP2 mixture.

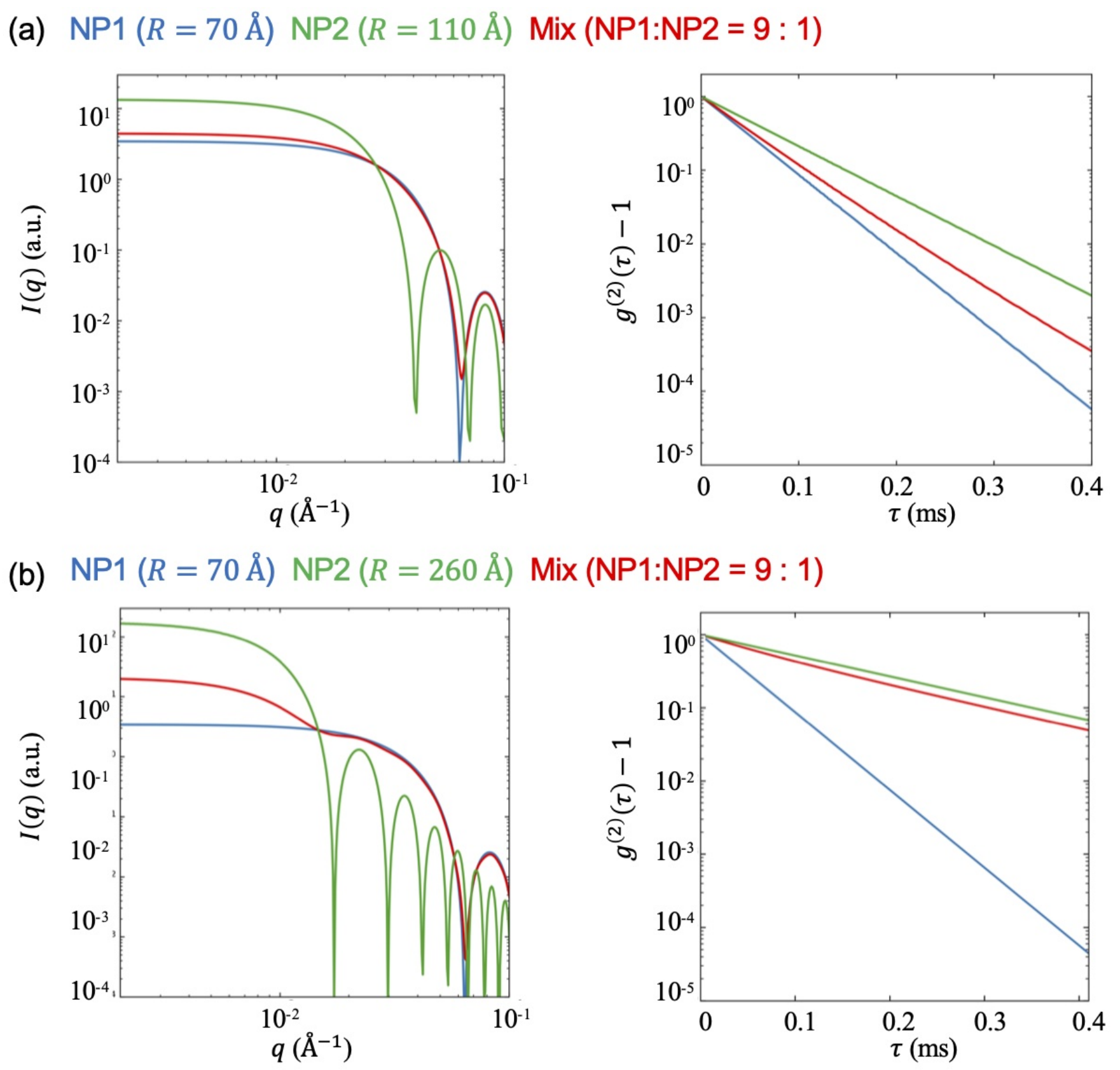

**Fig. 4 Simulation results for scattering intensity $I(q)$ and DLS correlation function.**

**a** $I(q)$ (left), calculated using Eq. (1), and $g^{(2)}(\tau) - 1$ (right), calculated using Eq. (2), for $R_1 = 70$ Å (blue),$R_2 = 110$ Å(green), and a mixture of NP1 and NP2 at a volume ratio of 9:1 (red).

**b** $I(q)$ (left), calculated using Eq. (1), and $g^{(2)}(\tau) - 1$ (right), calculated using Eq. (2), for $R_1 = 70$ Å (blue), $R_2 = 260$ Å (green), and a mixture of NP1 and NP2 at a volume ratio of 9:1 (red).